\documentclass{article}

\usepackage[final,nonatbib]{neurips_2022}

\usepackage{caption}
\usepackage{amsmath}
\usepackage{graphicx}
\usepackage[utf8]{inputenc} 
\usepackage[T1]{fontenc}    
\usepackage{hyperref}       
\usepackage{url}            
\usepackage{booktabs}       
\usepackage{amsfonts}       
\usepackage{nicefrac}       
\usepackage{microtype}      
\usepackage{xcolor}         
\usepackage{appendix}   
\usepackage{booktabs}  
\usepackage{caption}   
\usepackage{float}   
\usepackage{listings}
\usepackage{scrextend}
\usepackage{longtable}

\usepackage[style=numeric-comp]{biblatex}
\title{Large Language Model Prediction Capabilities: \\ Evidence from a Real-World 
 Forecasting Tournament}

\author{%
 Philipp Schoenegger \\
  London School of Economics\\and Political Science\\
   \And
   Peter S. Park \\
  MIT \\
}

\begin{document}

\maketitle

\begin{abstract}
Accurately predicting the future would be an important milestone in the capabilities of artificial intelligence. However, research on the ability of large language models to provide probabilistic predictions about future events remains nascent. To empirically test this ability, we enrolled OpenAI's state-of-the-art large language model, GPT-4, in a three-month forecasting tournament hosted on the Metaculus platform. The tournament, running from July to October 2023, attracted 843 participants and covered diverse topics including Big Tech, U.S. politics, viral outbreaks, and the Ukraine conflict. Focusing on binary forecasts, we show that GPT-4's probabilistic forecasts are significantly less accurate than the median human-crowd forecasts. We find that GPT-4's forecasts did not significantly differ from the no-information forecasting strategy of assigning a 50\% probability to every question. We explore a potential explanation, that GPT-4 might be predisposed to predict probabilities close to the midpoint of the scale, but our data do not support this hypothesis. Overall, we find that GPT-4 significantly underperforms in real-world predictive tasks compared to median human-crowd forecasts. A potential explanation for this underperformance is that in real-world forecasting tournaments, the true answers are genuinely unknown at the time of prediction; unlike in other benchmark tasks like professional exams or time series forecasting, where strong performance may at least partly be due to the answers being memorized from the training data. This makes real-world forecasting tournaments an ideal environment for testing the generalized reasoning and prediction capabilities of artificial intelligence going forward.
\end{abstract}

\section{Introduction}

In the field of artificial intelligence (AI), large language models (LLMs) have recently shown surprising capabilities in a multitude of economically relevant tasks \supercite{naveed2023comprehensive} that were previously thought to require human cognition.\supercite{bubeck2023sparks} State-of-the-art LLMs are comprised of an extremely large amount of parameters---typically organized in terms of the Transformer architecture\supercite{vaswani2017attention}---and trained on a large corpus of Internet-based text data. This corpus of training data is used to train LLMs to predict the next sequence of tokens given an input. Despite the simplicity of the general task for which LLMs are trained---next token prediction---the resulting transfer learning causes LLMs to also become ostensibly proficient at a wide variety of specific tasks, including reading comprehension, \supercite{deWinterJoostC, park2023diminished} summarization, \supercite{goyal2023news} translation, \supercite{jiao2023chatgpt} coding, \supercite{bubeck2023sparks} deception, \supercite{park2023ai} medical-license exams \supercite{nori2023capabilities,bubeck2023sparks}, and bar exams. \supercite{katz2023gpt,bubeck2023sparks} 

The ostensibly impressive capabilities of LLMs come with several important caveats. One of the most important caveats to keep in mind is that when an LLM consistently outputs the true answers to questions in a task benchmark or exam, it is unclear whether the LLM’s true answers reflect a genuine understanding of the task: and as such, are likely to generalize to out-of-distribution settings.\supercite{arora2023theory} An alternative hypothesis is that the training dataset contains at least part of the task benchmark’s questions and corresponding answers, and that the given LLM capability may thus not reflect a genuine understanding that can generalize to analogous questions not in the training dataset.\supercite{stochasticparrots,magar-schwartz-2022-data,CarliniIJLTZ23,biderman2023emergent} For example, consider a scenario where an organic-chemistry exam question is inputted into the LLM, and it outputs the true answer. It remains unclear whether this is due to the LLM possessing a deep understanding of the relevant organic-chemistry concepts, as opposed to simply reproducing the answer to the specific question contained in its training dataset. It is not trivial to rigorously formalize the dichotomy between genuine understanding and training-data memorization, as genuine understanding also ultimately arises from the relevant content in the training dataset. But the very real phenomenon of generalizability---or lack thereof---seems to be at the heart of the dichotomy.\supercite{grove2012continuum}

Given the problem of distinguishing whether an LLM's proficiency at a task benchmark is due to genuine understanding or to memorizing the training data, we propose evaluating LLM capabilities in contexts where the true answers are unknown beforehand and as such cannot be part of the training data. An especially important such context is forecasting, a domain where the answers are initially unknown to everyone (even the human evaluators) until they are validated or falsified by future events.\supercite{petropoulos2022forecasting}  We distinguish this context from the related but different forecasting context of zero-shot extrapolations of time series.\supercite{gruver2023large} In contrast, our definition of `forecasting'---producing accurate probabilistic predictions of future events---does not encounter the problem of distinguishing the benchmark data from the training data. As such, it allows for a more robust and generalizable evaluation of LLM capabilities: beyond the rote memorization of training data.

Our study evaluates the forecasting capabilities of GPT-4, a state-of-the-art LLM created by OpenAI,\supercite{openai2023gpt4} by entering it in a forecasting tournament. Forecasting tournaments are competitions wherein individual forecasters provide probabilistic forecasts on questions concerning future events. \supercite{tetlock2014forecasting}  These predictions are then scored based on the accuracy of their forecasts: the closer one’s prediction is to the truth, the more likely one is to be rewarded. The collective accuracy of the predictions resulting from such forecasting tournaments hinges on the 'wisdom of the crowd,'  i.e., the observation that aggregated forecasts of a group are often more accurate than the forecasts of individuals, \supercite{mannes2014wisdom,budescu2015identifying} even if judgments are correlated and biased.\supercite{davis2014crowd} This setting allows us to test LLM capabilities in a context where the answer is genuinely unknown beforehand and thus cannot be contained within the training dataset. This setting may also be critical for whether LLMs are or will be able to rival---or even outperform---humans at jobs that require prescient decision-making.\supercite{park2023divideandconquer} Accurately predicting the future is foundational to decision-making in most public and private sectors.\supercite{petropoulos2022forecasting} Consequently, comparing the forecasting capabilities of GPT-4 with those of humans promises to be a good test case for many potential economic applications of advanced AI.

As such, we have pre-registered---along with our analysis plans---the following null hypothesis: 
\begin{quote}
The mean accuracy of GPT-4 forecasts is not different from the mean accuracy of median human-crowd forecasts. \(H_0: \mu_{\text{GPT-4}} = \mu_{\text{Human}}\)
\end{quote}

Human crowds in forecasting tournaments are among the best options for producing accurate probabilistic predictions.\supercite{tetlock2017bringing,dardaman2023asking}If a state-of-the-art LLM like GPT-4 is able to outperform a human crowd of forecasters, this would be consistent with the model having learned a deep understanding of the relevant capabilities, like probabilistic reasoning, generalization, and accurate prediction.

 \section{Methods}
We conducted our study on Metaculus, a platform that hosts forecasting tournaments where members of the public can submit predictions, compete for prizes, and establish a forecasting track record. Metaculus' forecasting platform has been employed in various academic and policy prediction contexts, such as the monkeypox outbreak in 2022 \supercite{mcandrew2022early} and the COVID pandemic.\supercite{mcandrew2022chimeric} We leveraged the opportunity provided by the launch of the Quarterly Cup\supercite{metaculus2023} on July 3, 2023, as this three-month tournament offered an ideal context to evaluate LLM prediction capabilities. This setting provided us with a context where an LLM could compete with human forecasters to make probabilistic predictions about various topical questions.

Our questions were vetted by the Metaculus moderation team, who aimed to resolve the questions in a consistent and accurate manner in order to ensure high-quality data. The forecasting questions studied here were on a wide array of topics, such as U.S. industrial action disputes, military interventions in Niger, outbreaks of Marburg virus disease, and the Black Sea grain deal. For examples of the forecasting questions used in our study, see Table~\ref{tab:questions}. For a full set of questions, see \ref{sec:forecasting_questions}.

\begin{table}[h!]
\centering
\captionsetup{skip=10pt}  
\begin{tabular}{p{0.9\textwidth}}
\toprule
Will the United Auto Workers call a strike against any of the Big Three Detroit automakers before September 19, 2023? \\
\midrule
Will Mohamed Bazoum, Nigerien President, return to power before August 31, 2023?\\
\midrule
Will India's Chandrayaan-3 mission successfully land a rover on the moon? \\
\midrule
Will the Black Sea grain deal be revived before October 1, 2023? \\
\midrule
Will a non-proprietary LLM be in the top 5 of the chat.lmsys.org leaderboard on September 30, 2023? \\
\bottomrule
\end{tabular}
\caption{Examples of forecasting questions used in our study.}
\label{tab:questions}
\end{table}

Our study's questions were answered by both GPT-4 and a large set of human forecasters. The tournament concluded on October 4, with a total of 51 questions posed, and with 843 unique forecasters entering at least one prediction.  For our analysis, we focused solely on the subset of binary questions, of which there were 23. This is because binary questions were the only types of questions that our LLM forecaster could straightforwardly answer without requiring additional human input. This could have biased the predictions, such as by drawing distributions based on quartile point estimates.

	We used the web interface of GPT-4 at the default temperature value. Temperature is a hyperparameter that controls the randomness of a model's output, with higher temperature settings resulting in more random outputs, and lower temperature settings resulting in more deterministic outputs.  Our prompt was crafted by drawing on established research in forecasting, with the aim of steering GPT-4 towards comfortably providing numerical predictions and of enhancing the accuracy of these predictions. First, we prompted the model to emulate a superforecaster,\supercite{tetlock2014forecasting} aligning with the best-practice recommendations of how to prompt models to act as if they were domain experts. \supercite{xu2023expertprompting} This part of the prompt enabled us to consistently avoid getting responses of a different format than explicit probabilistic forecasts. 

Second, we grounded our prompt on research from the forecasting literature that highly complex qualitative rationales and the use of base rates are associated with forecasting accuracy.\supercite{karvetski2022forecasting} This part of the prompt was an attempt to not only elicit probabilistic forecasts, but to ensure that these are as highly accurate as possible. 

The overall prompt is given by the following. 
\begin{quote}
	\textbf{Prompt}: In this chat, you are a superforecaster that has a strong track record of accurate forecasts of the future. 	As an experienced forecaster, you evaluate past data and trends carefully and aim to predict future events as 	accurately as you can, even though you cannot know the answer. This means you put probabilities on outcomes that you are uncertain about (ranging from 0 to 100\%). When the outcome is 	continuous, you give me 25th interquartile ranges. You also quickly outline your rationale. In your rationales, you carefully consider the reasons for and against your probability estimate, you will make use of comparison classes of similar events and probabilities and take into account base rates and past events as well as other forecasts and predictions. You will also consider different perspectives.
 \end{quote}

Each question of the tournament was predicted via a single chat log, with the prompt positioned at the outset. Following this, GPT-4 was supplied with the background information, the specific resolution criteria of the question, and the final question text exactly as presented on Metaculus. This information, provided by Metaculus users or staff, helped mitigate the temporal distance between the training data set cutoff data and the question launch, which is less fundamental of an information gap for the human comparison group. For a full set of this information for one question, see \ref{sec:additional_details_other}. For some questions, the  median human-crowd prediction was available at the time of forecast; while for others, it was not. In the case of the latter, we included the median human-crowd prediction in the description provided to GPT-4. 

Subsequently, we collected the probabilistic prediction and inputted it into the question, in percent form. In cases where the model output was a probability range, the mean probability of the two values was used instead. We abstained from making any revisions to the initial forecasts. We did so because determining when to update and what information to employ for this update would have necessitated a significant degree of human involvement in the process, which we sought to minimize given that updating is a crucial skill for human forecasters.\supercite{mellers2015identifying} Since repeated updating is fundamental to the method of crowd-aggregated forecasting, it was important to account for GPT-4's inability to autonomously carry out this task in order to enable unbiased inferences from our results.

In order to control for these confounding factors, our design entailed recording both the initial model forecast and the crowd-aggregated median forecast at the close of the same day at which the model forecast was submitted. This allowed us to control for updating over extended question periods, which is done by humans, but is difficult to operationalize for GPT-4 without introducing bias in deciding when and how to prompt the update. Our comparison thus directly put GPT-4 on par with the human forecasters by only looking at initial forecasts. The full data encompassed for each question GPT-4's forecast, the median human-crowd forecast at the conclusion of the same day, the ground truth, whether a community prediction was visible at the time of prediction, the number of human forecasters at this point in time (for a median number of 37 forecasters), and the total duration, spanning the timeframe from when predictions were made to the eventual resolution of the question. This set of data was collected for all 23 binary forecasts.

\section{Results}
\begin{figure}[h]
\centering
\includegraphics[width=0.8\textwidth]{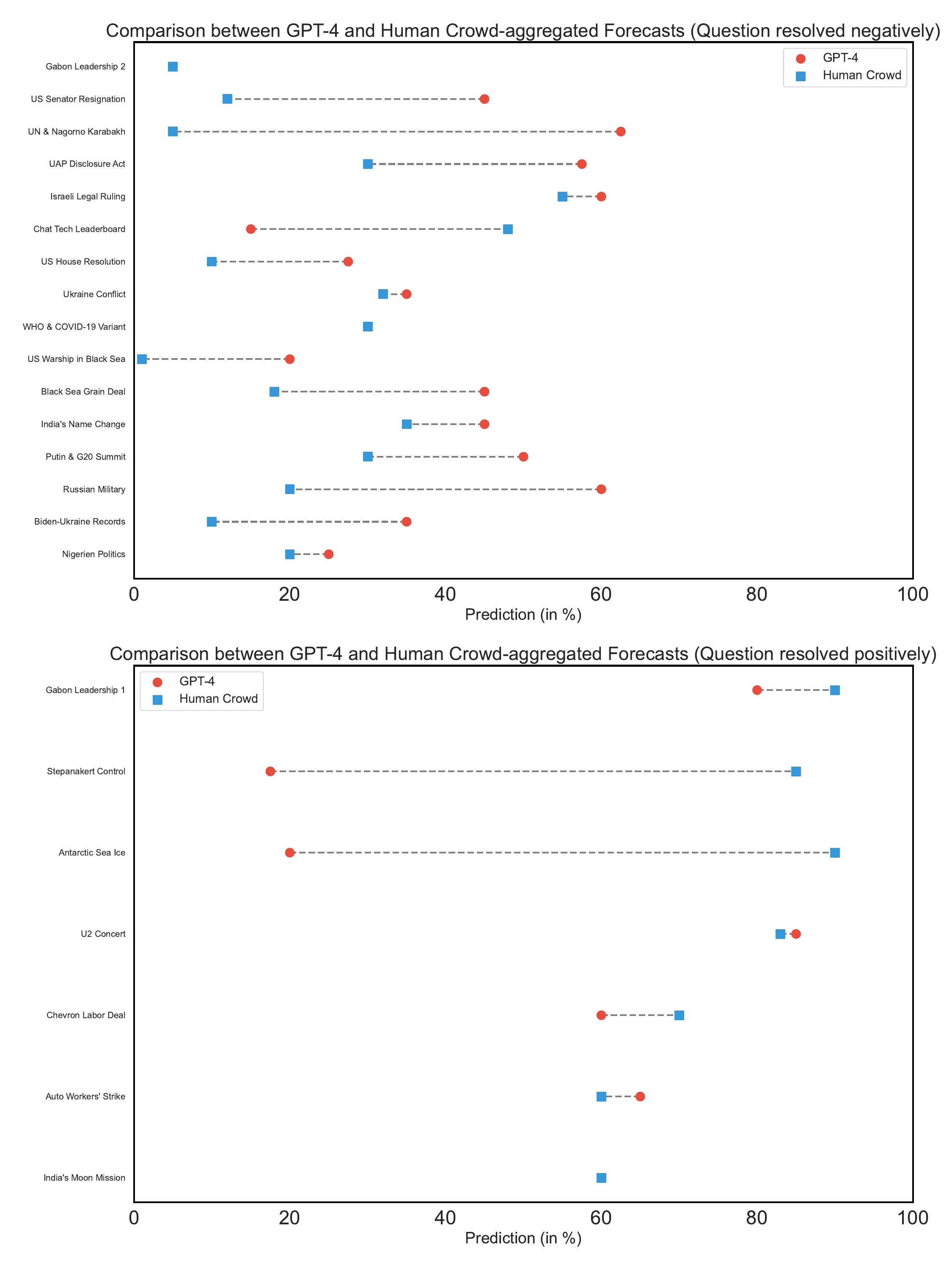}
\caption{Paired comparison plot showing probability forecasts (in \%) per question for the median human-crowd forecast and the GPT-4 forecast. The top panel lists questions that resolved negatively. The bottom panel lists questions that resolved positively.}\label{fig:1}
\end{figure}

First, we examine the probabilistic forecasts of both GPT-4 and the human crowd for each question in a directional analysis. See Figure~\ref{fig:1} for a paired comparison plot illustrating the respective probability forecasts, with the upper panel displaying questions that resolved positively and the lower panel displaying questions that resolved negatively. These data indicate that in 18 out of 23 questions, the median human-crowd forecasts were directionally closer to the truth than GPT-4's predictions, $\chi^2(1)=12.52, p=.001$. 

We also compared the frequency with which each of GPT-4's forecast and the aggregate human forecast was directionally correct, i.e., on the correct side of the probability midpoint. When doing so, we observed that the GPT-4 forecast was directionally correct 69.57\% of the time, in contrast to 95.65\% of the time for the aggregate human forecast, although the difference in distributions between the two was not statistically significant, $\chi^2(1)=3.78, p=.052$. Similarly, the 69.57\% proportion of directionally correct answers for GPT-4 also did not statistically deviate from the random 50\% baseline,  \( Z = 1.88, p = .061 \). These data suggest that while the aggregate human forecasts were comparatively more often closer to the truth than those of GPT-4, we do not find statistically significant results when testing this independently in relation to the probability scale midpoint. 

The primary pre-registered outcome of interest for our analysis is forecasting accuracy. To compute this accuracy as our dependent variable, we employ Brier scores,\supercite{brier1950verification} with the score for each individual provided below. Here, $f$ is the forecasted probability and $o$ is the observed outcome (which is 1 if the event occurs and 0 if it does not), given by $B=(f-o)^2$. We then aggregate the question-level Brier scores per condition as 
\begin{equation}
\frac{1}{N} \sum_{i=1}^N (f_{i,\text{condition}}-o_i)^2.
\end{equation}
 A perfect accuracy would yield a Brier score of 0, while perfect inaccuracy would result in a Brier score of 1.

In our data, we calculate the mean accuracy by aggregating the the question-level Brier scores for GPT-4's predictions and the median human-crowd forecasts. We observe an average Brier score for GPT-4's predictions of $B=.20$ ($SD = .18$), while the human forecaster average Brier score was $B=.07$ ($SD = .08$). Initially, we test both GPT-4's and the human crowd's accuracy against a simple no-information baseline. This baseline is the Brier score of 0.25, equivalent to predicting 50\% on each question. This serves as a first test of prediction accuracy. Our analysis reveals that GPT-4 does not exhibit predictive performance that is statistically distinct from the no-information baseline, $t(22)=-1.23, p=0.23$. On the other hand, we find that the human crowd's accuracy is significantly superior to that of the no-information baseline, $t(22)=-10.69, p<.001$. These findings suggest that while the aggregate human-crowd forecasts have high accuracy, the mean LLM forecasts do not significantly improve on the no-information baseline.

\begin{figure}[h]
\centering
\includegraphics[width=0.9\textwidth]{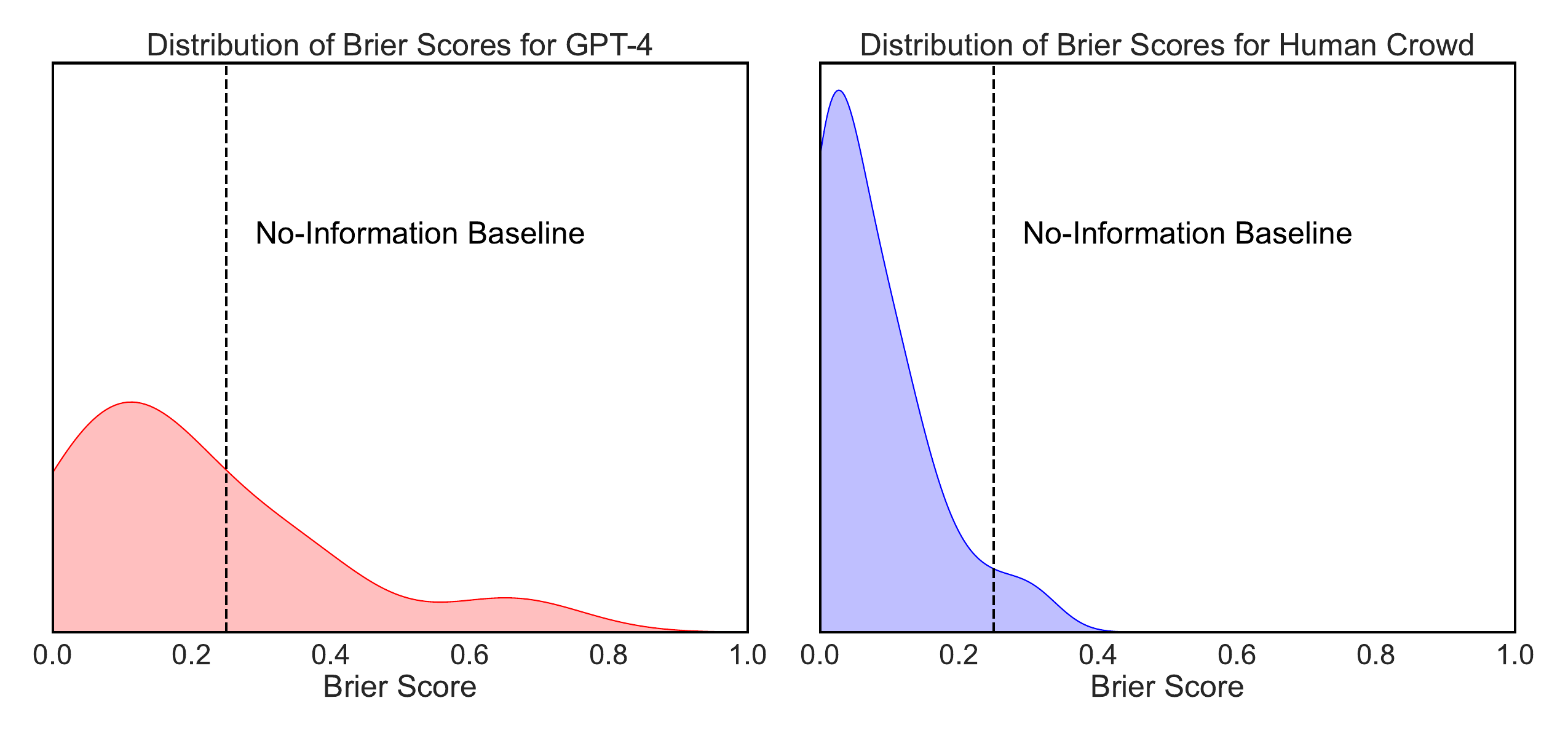}
\caption{Kernel Density Estimation (KDE) plots of Brier scores for GPT-4's forecasts and median human-crowd forecasts. Black dotted line represents a Brier score of 0.250 for the 'No-Information Baseline'.}\label{fig:2}
\end{figure}

In order to test our pre-registered null hypothesis, we compare the mean accuracy of GPT-4 and that of the human crowd. This yields a total mean difference of $\Delta=.13$, which is statistically significant at the two-tailed test, $t(44)=3.11,p=.003$, with an effect size of Cohen’s $d=.94$. Based on this, we reject our null hypothesis, and conclude that GPT-4 significantly underperforms in this real-world forecasting tournament compared to the median human-crowd forecasts that are currently employed in forecasting tournaments. We also find that GPT-4's accuracy was not significantly different on questions that included the community prediction compared to those that did not, $t(21)=-1.81, p=0.085$, suggesting that this result is explained by an internal factor of GPT-4's forecasting performance. For a graphical depiction of the distribution of Brier scores per condition, see Figure~\ref{fig:2}.

One potential reason that might explain this outcome could be that GPT-4's predictions, due to its reinforcement learning from human feedback,\supercite{ziegler2019fine} may gravitate towards the probability scale midpoint (50\%) in a form of model conservatism. This could explain the low prediction accuracy in our forecasting tournament with a small-to-medium sample size of questions. To analyse this, we conducted the following exploratory analyses. First, we computed the coefficient of variation (CV) as a normalized measure of dispersion for the predictions made by both GPT-4 and the human crowd. Our analysis found a CV of 48.22\% for GPT-4 and 73.67\% for the human crowd when centered on the mean, suggesting that GPT-4 has less dispersion. Similarly, when anchoring the CV around the midpoint of the probability scale, the values were 42.14\% for GPT-4 and 57.59\% for the human crowd respectively, again showing the pattern of GPT-4 having less dispersion.

\begin{figure}[h]
\centering
\includegraphics[width=0.9\textwidth]{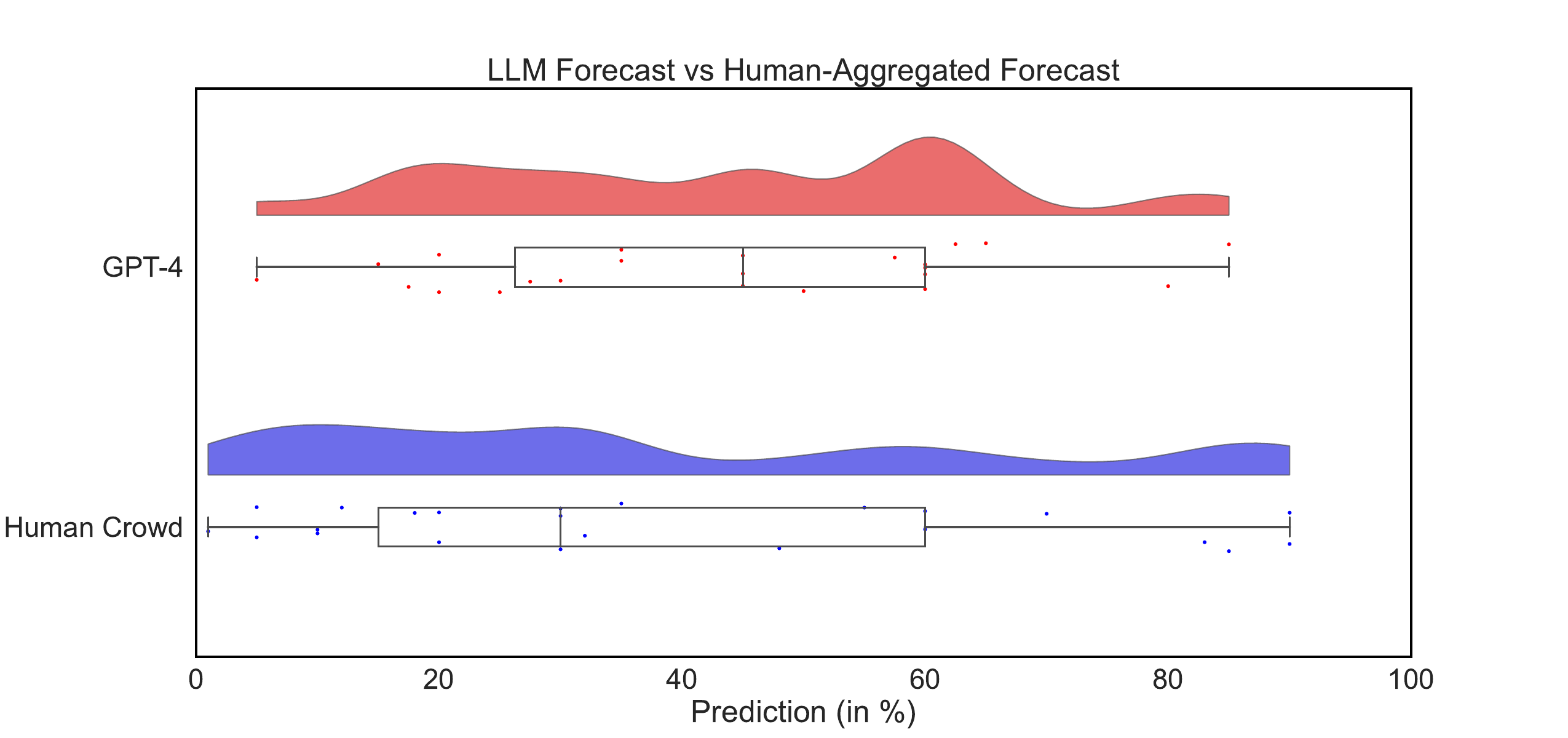}
\caption{Raincloud plot of the distribution of probability forecasts (in \%) made by GPT-4 and by the median human-crowd forecast for all questions. Box plots represent interquartile ranges. }\label{fig:3}
\end{figure}

These analyses might suggest that GPT-4's predictions could be more densely clustered around both the mean and the midpoint compared to the human crowd, aligning with a more cautious prediction pattern: one that is less inclined towards extreme forecasts. However, our inferential analysis employing Levene's test for equality of variances at the 50\% midpoint did not yield statistically significant evidence to substantiate differences in variance around the midpoint, $F(1,44)=.54, p=.47$. Thus, our results do not provide grounds for asserting a difference in variance around the probability scale midpoint. See Figure~\ref{fig:3} for a graphical representation of these findings.

\begin{figure}[h]
\centering
\includegraphics[width=0.9\textwidth]{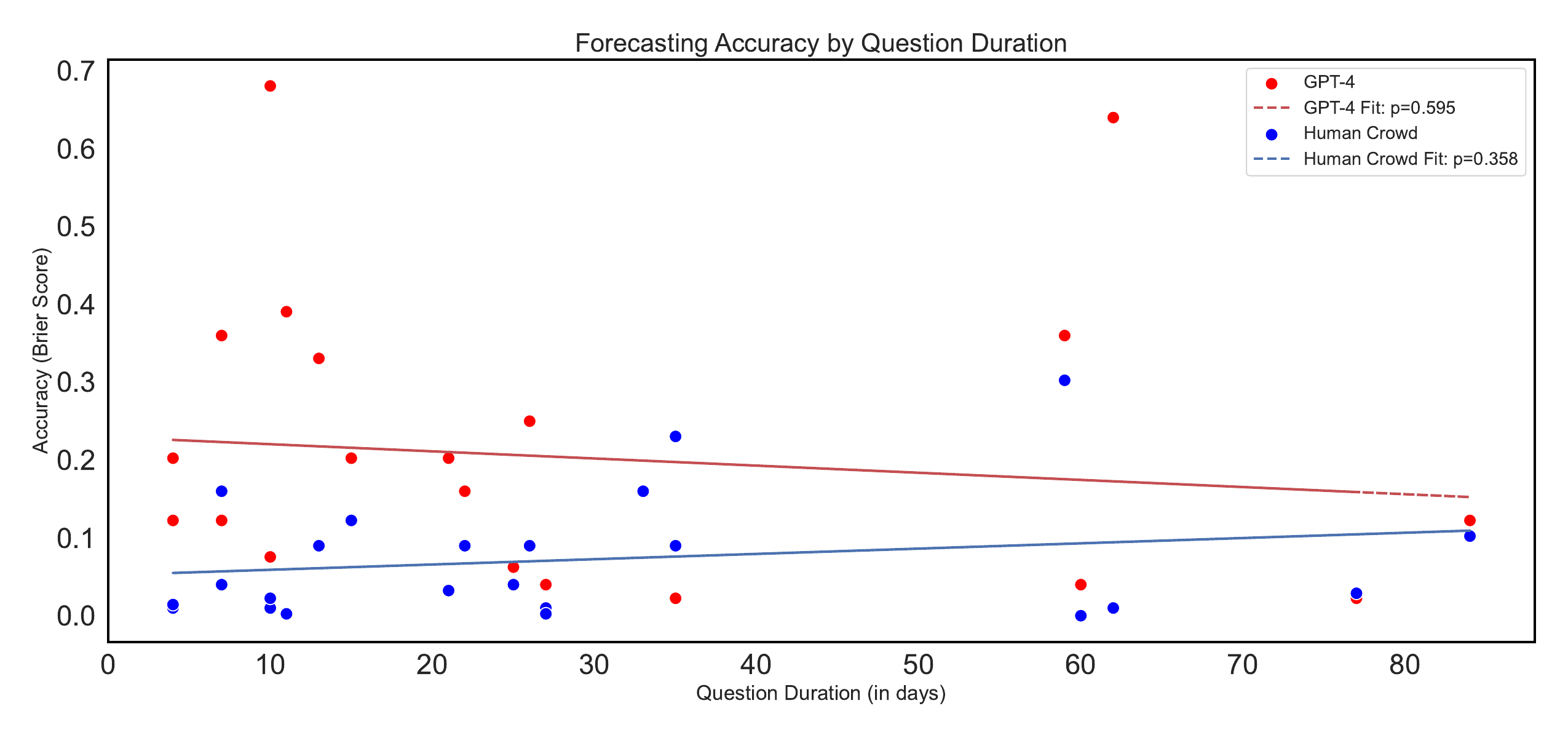}
\caption{Scatterplot and linear fit for the relationship between forecasting accuracy (Brier score) and duration (in days)  by condition.}\label{fig:4}
\end{figure}

Furthermore, in an additional exploratory analysis, we probe whether the relationship between the duration of a forecasting question remaining unresolved and the forecasting accuracy at the inception of this period significantly diverges between GPT-4's forecasts and human-crowd forecasts. This is interesting because it may be that GPT-4's forecasts are especially good, or especially bad, at predicting questions that resolve very quickly compared to those that do not. See Figure~\ref{fig:4} for a scatterplot illustrating accuracy and question duration with linear fits. Our results show that we do not detect statistically significant effects of duration, $ps >.36$. 

To test this question directly, we conducted an Analysis of Covariance (ANCOVA) to assess whether the relationship between accuracy and question duration differs between GPT-4's forecasts and those made by human forecasters. The main effect of the method (human crowd vs. GPT-4) was statistically significant, $F(1,42)=9.38,\textit{p}=.004$, indicating a notable difference in accuracy between the two forecasting methods. However, the main effect of duration was not significant,  $F(1,42)=0.017,\textit{p}=.898$, suggesting that question duration did not significantly affect the accuracy of forecasts. Moreover, the interaction between method and duration was also not significant, $F(1,42)=0.75,\textit{p}=.392$, suggesting that the relationship between question duration and accuracy did not significantly vary between GPT-4 and the human crowd.

Additionally, there are also a variety of aggregation techniques that may prove useful in combining the aggregate human forecast and GPT-4's forecast. We report the exploratory results of a Bayesian Model Averaging (BMA) approach, which takes into account the uncertainty across models prior to combining the predictions from multiple models. We first calculate the likelihoods of each model (.82 for GPT-4 and .93 for the human crowd) before normalizing them and computing the posterior model probabilities. With these, we have the following weighted average: 
\begin{equation}\text{Brier}_{\text{BMA}}=(0.467 \cdot \text{Brier}_{\text{GPT-4}} )+(0.533 \cdot \text{Brier}_{\text{Human Crowd}} ).
\end{equation}
We compute $\text{Brier}_{\text{BMA}}=.13$ ($SD=.09$), which is significantly more accurate than the random baseline, $t(22)=-5.91, p<.001$. 

We have presented this simple aggregation as an instance of how LLM predictions may feed into future aggregation approaches. We point out that given our finding of poor prediction accuracy, current models like GPT-4 may be unlikely to make worthwhile additions to aggregation algorithms. However, this may change with advances in LLM capabilities and LLM-forecasting technology.

\section{Discussion}
Our findings from entering GPT-4 into a real-world forecasting tournament on the Metaculus platform suggest that even this state-of-the-art LLM has unimpressive forecasting capabilities. Despite being prompted with established superforecasting techniques and best-practice prompting approaches, GPT-4 was heavily outperformed by the forecasts of the human crowd, and did not even outperform a no-information baseline of predicting 50\% on every question. The robustness of this finding is suggested by the fact that the question set was diverse, drawing on a wide variety of topics like Big Tech, U.S. politics, viral outbreaks, and the Ukraine conflict. We argue that our data also provide additional evidence for the predictive power of human crowd forecasting competitions, as their accuracy was impressive throughout the questions studied here. Our results thus suggest that current LLMs may not yet perform well in a variety of real-world prediction tasks that would necessitate probabilistic foresight. 

One potential contributing factor to GPT-4's inadequate forecasting capability is the fact that its training data is subject to a knowledge cutoff after a certain point in time.\supercite{openai2023gpt4} In contrast to human forecasters who keep up-to-date with recent events, GPT-4 is not updated with events that occur post-training. Given the dynamic nature of many world events, GPT-4's lack of real-time knowledge updating can be a significant barrier to accurate forecasting. Our design tried to address this concern by inputting the question's background information presented on Metaculus to the LLM, so that it would not be wholly unaware of a potentially novel context. While this does not fully bring the model up to speed with human forecasters that seek out prediction competition on Metaculus, it does help mitigate the information gap posed by the knowledge cutoff. Additional ways to mitigate the limitations to forecasting posed by LLMs' knowledge cutoff would be fruitful in future studies. \supercite{li2023largelanguagemodels}

OpenAI's mission is to create ``highly autonomous systems that outperform humans at most economically valuable work.''\supercite{openai2018charter} Whether this AI-led future is on track to occur will likely be in large part determined by how capable LLMs---and AI systems in general---turn out to be at economically relevant tasks,\supercite{park2023divideandconquer}   especially without human hand-holding. Forecasting is a task of especially high economic relevance, especially for a large proportion of white-collar fields---such as business, policy, and law---that rely heavily on accurate predictions across a myriad of contexts. As of now, our results suggest that even state-of-the-art AI systems are not yet posed to replace human expertise in these areas, due to the inadequate forecasting capabilities of these systems. However, it will be especially important to closely monitor AI capability advances in the domain of accurate forecasting, as foresight will remain an essential skill for autonomous systems.

Another relevant implication of AI forecasting capabilities---or lack thereof---is the risk arising from AI systems that are proficient at long-term planning. An AI system with robust long-term planning capabilities would be able to presciently pursue their goal, which can be a sizable and potentially catastrophic danger if the goal happens to be incompatible with the well-being of humans\supercite{ngo2023alignment} (e.g., the goal of engineering a pandemic that kills as many people as possible over the long run). Proficiency in long-term planning requires the ability to accurately forecast future scenarios. For many real-world tasks, such accurate forecasting is a highly complex endeavor, at which even many (but not all) humans arguably perform poorly.\supercite{tetlock2016superforecasting} Our finding that GPT-4 has particularly poor forecasting capabilities bolsters the case that the threat of an AI system planning in the long term against human interests remains presently low. However, this should not be taken as a reason to be complacent. The pace at which AI capabilities advance suggests that it remains crucial to continually monitor the progression of AI systems in terms of their forecasting abilities to ensure that their development remains robustly safe.

 Our results raise several avenues for further research. First, it may be argued that our findings might, at least in part, be explained by the human comparison group's ability to harness the wisdom of the crowd,\supercite{himmelstein2023wisdom} whereas the LLM prediction might be best understood as a single forecast only. One counterargument to this claim is that LLM predictions themselves may draw on a wisdom-of-the-crowd effect from their large and multifaceted training dataset. Future research on how to employ a larger ensemble of LLM forecasters that draw on diverse inputs, training datasets, prompts, temperature values, and other variations would be fruitful.

Second, there may be merit in implementing LLM forecasters that can access the internet for information and autonomously update their forecasts. While the design and implementation of such a setup pose challenges, particularly as most such design choices require human input which may introduce bias, it remains a noteworthy endeavor. Internet access may also help mitigate the design weakness outlined earlier, where the temporal distance between the training data cutoff and the forecasting question's start may impair contextual understanding of the question details. 

Third, we outlined a simple Bayesian model-averaging approach for aggregating human and machine forecasts. Future work may delve deeper into alternative ways to combine such forecasts, such as exploring ways of using LLM forecasts in extremizing algorithms or other ways of selectively choosing to favor or disfavor LLM forecasts depending on their eventual performance and comparative strengths and weaknesses. However, it is worth pointing out that given the current poor prediction performance, such approaches probably only become worthwhile once LLM foresight abilities are significantly improved. 

Finally, it would be productive to investigate the integration of human and LLM forecasts, such as by examining whether hybrid forecasting models outperform the standalone baselines. Such hybrid forecasters (i.e., human forecasters that rely upon LLM outputs to make predictions) may be able to combine the respective strengths of both human cognition and LLM cognition, which open up the potential for new forecasting techniques based on the augmentation---rather than replacement---of humans by AI.\supercite{benjamin2023hybrid}

Our data suggest a clear weakness in an otherwise impressive catalogue of LLM capabilities: predicting the future. This highlights a number of technical challenges and future research directions for how LLMs may be harnessed for forecasting, as well as for various real-world tasks that require or are helped by robust forecasting capabilities. Our results suggest that the wisdom of human crowds remains a powerful tool in providing probabilistic forecasts about the future. This context of a forecasting tournament may also prove to be an especially fruitful environment for probing the degree to which LLMs are capable of generalized reasoning and prediction.

\section*{Acknowledgements}

We thank Molly Hickman, Barbara Fasolo, Siti Liyana Azman, Indre Tuminauskaite, Jeffrey Haines, Isabel Juniewicz, Philip Tetlock, and the Metaculus team for helpful comments. P.S. is funded by the LSE Department of Management. P.S.P. is funded by the MIT Department of Physics.

\section*{Data availability}
Our pre-registration and analysis plan can be found in our Open Science Foundation database: \url{https://osf.io/tfbvd/?view_only=bfc6d61152654e3b84c36446f1858fb2}

\printbibliography

@article{deWinterJoostC,
issn = {1560-4292},
journal = {International Journal of Artificial Intelligence in Education},
year = {2023},
title = {Can ChatGPT Pass High School Exams on English Language Comprehension?},
author = {de Winter, Joost C. F.},
}

@misc{jiao2023chatgpt,
      title={Is ChatGPT a Good Translator? Yes with GPT-4 as the Engine}, 
      author={Wenxiang Jiao and Wenxuan Wang and Jen-tse Huang and Xing Wang and Zhaopeng Tu},
      year={2023},
      eprint={2301.08745},
      archivePrefix={arXiv},
      primaryClass={cs.CL}
}

@book{openai2018charter,
    title={OpenAI Charter},
    author={OpenAI},
publisher={OpenAI},
    year={2018},
    url={https://openai.com/charter}
}

@misc{openai2023gpt4,
      title={GPT-4 Technical Report}, 
      author={OpenAI},
      year={2023},
      eprint={2303.08774},
      archivePrefix={arXiv},
      primaryClass={cs.CL}
}

@misc{goyal2023news,
      title={News Summarization and Evaluation in the Era of GPT-3}, 
      author={Tanya Goyal and Junyi Jessy Li and Greg Durrett},
      year={2023},
      eprint={2209.12356},
      archivePrefix={arXiv},
      primaryClass={cs.CL}
}

@misc{bubeck2023sparks,
      title={Sparks of Artificial General Intelligence: Early Experiments with GPT-4}, 
      author={Sébastien Bubeck and Varun Chandrasekaran and Ronen Eldan and Johannes Gehrke and Eric Horvitz and Ece Kamar and Peter Lee and Yin Tat Lee and Yuanzhi Li and Scott Lundberg and Harsha Nori and Hamid Palangi and Marco Tulio Ribeiro and Yi Zhang},
      year={2023},
      eprint={2303.12712},
      archivePrefix={arXiv},
      primaryClass={cs.CL}
}

@article{katz2023gpt,
  title={GPT-4 Passes the Bar Exam},
  author={Katz, Daniel Martin and Bommarito, Michael James and Gao, Shang and Arredondo, Pablo},
  journal={SSRN},
  year={2023}
}

@misc{nori2023capabilities,
      title={Capabilities of GPT-4 on Medical Challenge Problems}, 
      author={Harsha Nori and Nicholas King and Scott Mayer McKinney and Dean Carignan and Eric Horvitz},
      year={2023},
      eprint={2303.13375},
      archivePrefix={arXiv},
      primaryClass={cs.CL}
}

@article{vaswani2017attention,
  title={Attention is All You Need},
  author={Vaswani, Ashish and Shazeer, Noam and Parmar, Niki and Uszkoreit, Jakob and Jones, Llion and Gomez, Aidan N and Kaiser, {\L}ukasz and Polosukhin, Illia},
  journal={Advances in Neural Information Processing Systems},
  volume={30},
  year={2017}
}

@misc{park2023ai,
      title={AI Deception: A Survey of Examples, Risks, and Potential Solutions}, 
      author={Peter S. Park and Simon Goldstein and Aidan O'Gara and Michael Chen and Dan Hendrycks},
      year={2023},
      eprint={2308.14752},
      archivePrefix={arXiv},
      primaryClass={cs.CY}
}

@misc{park2023diminished,
      title={Diminished Diversity-of-Thought in a Standard Large Language Model}, 
      author={Peter S. Park and Philipp Schoenegger and Chongyang Zhu},
      year={2023},
      eprint={2302.07267},
      archivePrefix={arXiv},
      primaryClass={cs.HC}
}

@misc{biderman2023emergent,
      title={Emergent and Predictable Memorization in Large Language Models}, 
      author={Stella Biderman and USVSN Sai Prashanth and Lintang Sutawika and Hailey Schoelkopf and Quentin Anthony and Shivanshu Purohit and Edward Raff},
      year={2023},
      eprint={2304.11158},
      archivePrefix={arXiv},
      primaryClass={cs.CL}
}

@inproceedings{CarliniIJLTZ23,
  author       = {Nicholas Carlini and
                  Daphne Ippolito and
                  Matthew Jagielski and
                  Katherine Lee and
                  Florian Tram{\`{e}}r and
                  Chiyuan Zhang},
  title        = {Quantifying Memorization Across Neural Language Models},
  booktitle    = {The Eleventh International Conference on Learning Representations,
                  {ICLR} 2023, Kigali, Rwanda, May 1-5, 2023},
  publisher    = {OpenReview.net},
  year         = {2023},
  url          = {https://openreview.net/pdf?id=TatRHT\_1cK},
  bibsource    = {dblp computer science bibliography, https://dblp.org}
}

@article{arora2023theory,
  title={A Theory for Emergence of Complex Skills in Language Models},
  author={Arora, Sanjeev and Goyal, Anirudh},
  journal={arXiv preprint arXiv:2307.15936},
  year={2023}
}

@misc{ngo2023alignment,
      title={The Alignment Problem from a Deep Learning Perspective}, 
      author={Richard Ngo and Lawrence Chan and Sören Mindermann},
      year={2023},
      eprint={2209.00626},
      archivePrefix={arXiv},
      primaryClass={cs.AI}
}

@inproceedings{magar-schwartz-2022-data,
    title = "Data Contamination: From Memorization to Exploitation",
    author = "Magar, Inbal  and
      Schwartz, Roy",
    booktitle = "Proceedings of the 60th Annual Meeting of the Association for Computational Linguistics (Volume 2: Short Papers)",
    month = may,
    year = "2022",
    address = "Dublin, Ireland",
    publisher = "Association for Computational Linguistics",
    url = "https://aclanthology.org/2022.acl-short.18",
    doi = "10.18653/v1/2022.acl-short.18",
    pages = "157--165",
}

@inproceedings{dardaman2023asking,
    author = {Dardaman, Emily and Gupta, Abhishek},
    title = {Asking Better Questions: The Art and Science of Forecasting},
    booktitle = {CHI 2023 Designing Technology and Policy Simultaneously: Towards A Research Agenda and New Practice Workshop},
    year = {2023},
    address = {Hamburg, Germany},
    publisher = {ACM},
    url = {https://doi.org/...}  % Replace with the actual DOI or URL, if available
}

@article{tetlock2014forecasting,
  title={Forecasting Tournaments: Tools for Increasing Transparency and Improving the Quality of Debate},
  author={Tetlock, Philip E. and Mellers, Barbara A and Rohrbaugh, Nick and Chen, Eva},
  journal={Current Directions in Psychological Science},
  volume={23},
  number={4},
  pages={290--295},
  year={2014},
  publisher={Sage Publications Sage CA: Los Angeles, CA}
}

@article{mcandrew2022early,
  title={Early Human Judgment Forecasts of Human Monkeypox, May 2022},
  author={McAndrew, Thomas and Majumder, Maimuna S and Lover, Andrew A and Venkatramanan, Srini and Bocchini, Paolo and Besiroglu, Tamay and Codi, Allison and Braun, David and Dempsey, Gaia and Abbott, Sam and others},
  journal={The Lancet Digital Health},
  volume={4},
  number={8},
  pages={e569--e571},
  year={2022},
  publisher={Elsevier}
}

@online{metaculus2023,
    author    = {Metaculus},
    title     = {Quarterly Cup},
    year      = {2023},
publisher={Metaculus},
    url       = {https://www.metaculus.com/tournament/quarterly-cup-2023q3/}
}

@article{ziegler2019fine,
  title={Fine-tuning Language Models from Human Preferences},
  author={Ziegler, Daniel M and Stiennon, Nisan and Wu, Jeffrey and Brown, Tom B and Radford, Alec and Amodei, Dario and Christiano, Paul and Irving, Geoffrey},
  journal={arXiv preprint arXiv:1909.08593},
  year={2019}
}

@misc{gruver2023large,
      title={Large Language Models Are Zero-Shot Time Series Forecasters}, 
      author={Nate Gruver and Marc Finzi and Shikai Qiu and Andrew Gordon Wilson},
      year={2023},
      eprint={2310.07820},
      archivePrefix={arXiv},
      primaryClass={cs.LG}
}

@article{mcandrew2022chimeric,
  title={Chimeric Forecasting: Combining Probabilistic Predictions from Computational Models and Human Judgment},
  author={McAndrew, Thomas and Codi, Allison and Cambeiro, Juan and Besiroglu, Tamay and Braun, David and Chen, Eva and De C{\`e}saris, Luis Enrique Urtubey and Luk, Damon},
  journal={BMC Infectious Diseases},
  volume={22},
  number={1},
  pages={833},
  year={2022},
  publisher={Springer}
}

@misc{park2023divideandconquer,
      title={Divide-and-Conquer Dynamics in AI-Driven Disempowerment}, 
      author={Peter S. Park and Max Tegmark},
      year={2023},
      eprint={2310.06009},
      archivePrefix={arXiv},
      primaryClass={cs.CY}
}

@article{grove2012continuum,
  title={A Continuum of Learning: From Rote Memorization to Meaningful Learning in Organic Chemistry},
  author={Grove, Nathaniel P and Bretz, Stacey Lowery},
  journal={Chemistry Education Research and Practice},
  volume={13},
  number={3},
  pages={201--208},
  year={2012},
  publisher={Royal Society of Chemistry}
}

@misc{naveed2023comprehensive,
      title={A Comprehensive Overview of Large Language Models}, 
      author={Humza Naveed and Asad Ullah Khan and Shi Qiu and Muhammad Saqib and Saeed Anwar and Muhammad Usman and Nick Barnes and Ajmal Mian},
      year={2023},
      howpublished={\url{https://github.com/humza909/LLM_Survey.git}},
}

@inproceedings{li2023largelanguagemodels,
      title={Large Language Models with Controllable Working Memory},
      author={Daliang Li and Ankit Singh Rawat and Manzil Zaheer and Xin Wang and Michal Lukasik and Andreas Veit and Felix Yu and Sanjiv Kumar},
      booktitle={Findings of the Association for Computational Linguistics: ACL 2023},
      pages={1774--1793},
      year={2023},
      address={Toronto, Canada},
      publisher={Association for Computational Linguistics},
      url={https://aclanthology.org/2023.findings-acl.112},
      doi={10.18653/v1/2023.findings-acl.112}
}

@incollection{himmelstein2023wisdom,
  title={The Wisdom of Timely Crowds},
  author={Himmelstein, Mark and Budescu, David V. and Han, Ying},
  booktitle={Judgment in Predictive Analytics},
  pages={215--242},
  year={2023},
  publisher={Springer}
}

@article{mellers2015identifying,
  title={Identifying and Cultivating Superforecasters as a Method of Improving Probabilistic Predictions},
  author={Mellers, Barbara and Stone, Eric and Murray, Terry and Minster, Angela and Rohrbaugh, Nick and Bishop, Michael and Chen, Eva and Baker, Joshua and Hou, Yuan and Horowitz, Michael and others},
  journal={Perspectives on Psychological Science},
  volume={10},
  number={3},
  pages={267--281},
  year={2015},
  publisher={Sage Publications Sage CA: Los Angeles, CA}
}

@article{karvetski2022forecasting,
  title={What do Forecasting Rationales Reveal about Thinking Patterns of Top Geopolitical Forecasters?},
  author={Karvetski, Christopher W and Meinel, Carolyn and Maxwell, Daniel T and Lu, Yunzi and Mellers, Barbara A and Tetlock, Philip E.},
  journal={International Journal of Forecasting},
  volume={38},
  number={2},
  pages={688--704},
  year={2022},
  publisher={Elsevier}
}

@article{davis2014crowd,
  title={When is a Crowd Wise?},
  author={Davis-Stober, Clintin P. and Budescu, David V. and Dana, Jason and Broomell, Stephen B.},
  journal={Decision},
  volume={1},
  number={2},
  pages={79},
  year={2014},
  publisher={Educational Publishing Foundation}
}

@misc{xu2023expertprompting,
      title={ExpertPrompting: Instructing Large Language Models to be Distinguished Experts}, 
      author={Benfeng Xu and An Yang and Junyang Lin and Quan Wang and Chang Zhou and Yongdong Zhang and Zhendong Mao},
      year={2023},
      eprint={2305.14688},
      archivePrefix={arXiv},
      primaryClass={cs.CL}
}

@article{budescu2015identifying,
  title={Identifying Expertise to Extract the Wisdom of Crowds},
  author={Budescu, David V and Chen, Eva},
  journal={Management Science},
  volume={61},
  number={2},
  pages={267--280},
  year={2015},
  publisher={INFORMS}
}

@article{tetlock2017bringing,
  title={Bringing Probability Judgments into Policy Debates via Forecasting Tournaments},
  author={Tetlock, Philip E. and Mellers, Barbara A and Scoblic, J Peter},
  journal={Science},
  volume={355},
  number={6324},
  pages={481--483},
  year={2017},
  publisher={American Association for the Advancement of Science}
}

@article{mannes2014wisdom,
  title={The Wisdom of Select Crowds},
  author={Mannes, Albert E and Soll, Jack B and Larrick, Richard P},
  journal={Journal of Personality and Social Psychology},
  volume={107},
  number={2},
  pages={276},
  year={2014},
  publisher={American Psychological Association}
}

@article{petropoulos2022forecasting,
  title={Forecasting: Theory and Practice},
  author={Petropoulos, Fotios and Apiletti, Daniele and Assimakopoulos, Vassilios and Babai, Mohamed Zied and Barrow, Devon K and Taieb, Souhaib Ben and Bergmeir, Christoph and Bessa, Ricardo J and Bijak, Jakub and Boylan, John E and others},
  journal={International Journal of Forecasting},
  volume={38},
  number={3},
  pages={705--871},
  year={2022},
  publisher={Elsevier}
}

@book{tetlock2016superforecasting,
  title={Superforecasting: The Art and Science of Prediction},
  author={Tetlock, Philip E. and Gardner, Dan},
  year={2016},
  publisher={Random House}
}

@article{brier1950verification,
  title={Verification of Forecasts Expressed in Terms of Probability},
  author={Brier, Glenn W},
  journal={Monthly Weather Review},
  volume={78},
  number={1},
  pages={1--3},
  year={1950},
  publisher={American Meteorological Society}
}

@article{benjamin2023hybrid,
  title={Hybrid Forecasting of Geopolitical Events},
  author={Benjamin, Daniel M. and Morstatter, Fred and Abbas, Ali E. and Abeliuk, Andres and Atanasov, Pavel and Bennett, Stephen and Beger, Andreas and Birari, Swapnil and Budescu, David V. and Catasta, Michele and Ferrara, Emilio},
  journal={AI Magazine},
  year={2023}
}

@inproceedings{stochasticparrots,
author = {Bender, Emily M. and Gebru, Timnit and McMillan-Major, Angelina and Shmitchell, Shmargaret},
title = {On the Dangers of Stochastic Parrots: Can Language Models be too Big?},
year = {2021},
isbn = {9781450383097},
publisher = {Association for Computing Machinery},
address = {New York, NY, USA},
doi = {10.1145/3442188.3445922},
booktitle = {Proceedings of the 2021 ACM Conference on Fairness, Accountability, and Transparency},
pages = {610–623},
numpages = {14},
location = {Virtual Event, Canada},
series = {FAccT '21}
}

\newpage

\appendix
\renewcommand{\thesection}{Appendix \Alph{section}} 

\section{Forecasting questions}  
\label{sec:forecasting_questions}

\begin{table}[h]
\centering
\captionsetup{skip=10pt}  
\begin{tabular}{p{0.9\textwidth}}
\toprule
Will India's Chandrayaan-3 mission successfully land a rover on the moon? \\
\midrule
Will Mohamed Bazoum, Nigerien President, return to power before August 31, 2023? \\
\midrule
Will the House Oversight Committee receive access to requested Joe Biden records related to Hunter Biden’s Ukraine dealings before September 1st? \\
\midrule
Will General Sergei Surovikin be stripped of his command by July 11th? \\
\midrule
Will Putin attend the G20 summit in India? \\
\midrule
Will the United Auto Workers call a strike against any of the Big Three Detroit automakers before September 19, 2023? \\
\midrule
Will Chevron reach an agreement with the Offshore Alliance to end or prevent industrial actions before September 25, 2023? \\
\midrule
Will a bill be introduced in the Indian Parliament to change the official name of the country to Bharat before September 23, 2023? \\
\midrule
Will the Black Sea grain deal be revived before October 1, 2023? \\
\midrule
Will a US warship enter the Black Sea before September 25, 2023? \\
\midrule
Will the U2 concert at The Sphere on September 29, 2023 take place? \\
\midrule
Will the WHO name BA.2.86 as a SARS-CoV-2 Variant of Interest before October 1, 2023? \\
\midrule
Will the extent of Antarctic sea ice for every day in September 2023 be the lowest in recorded history? \\
\midrule
Will Ukraine regain control of central Bakhmut by the end of September 2023? \\
\midrule
Will a vote on a Republican-introduced resolution to vacate the Speaker of the House be held before October 1, 2023? \\
\midrule
Will a non-proprietary LLM be in the top 5 of the chat.lmsys.org leaderboard on September 30, 2023? \\
\midrule
Will the Israeli High Court issue a ruling on the 'reasonableness' law before October 1, 2023? \\
\midrule
Will the US pass the Unidentified Anomalous Phenomena (UAP) Disclosure Act of 2023 before October 1, 2023? \\
\midrule
Will the UN Security Council adopt a resolution related to the Nagorno Karabakh conflict between Azerbaijan and Armenia before October 1, 2023? \\
\midrule
Will Stepanakert / Khankendi be under de facto Azerbaijani control on September 30, 2023? \\
\midrule
Before October 1, 2023, will US Senator Bob Menendez announce that he is resigning? \\
\midrule
Who will be the de facto leader of Gabon on September 30, 2023 (General Brice Oligui Nguema)? \\
\midrule
Who will be the de facto leader of Gabon on September 30, 2023? (Albert Ondo Ossa)\\
\bottomrule
\end{tabular}
\caption{Full list of forecasting questions used in our study.}
\label{tab:questions_appendix}
\end{table}

\section{Example forecasting question and its corresponding information}  
\label{sec:additional_details_other}

\subsection*{Question}
Will a vote on a Republican-introduced resolution to vacate the Speaker of the House be held before October 1, 2023?

\subsection*{Resolution Criteria}
This question resolves as Yes if, before October 1, 2023, a member of the Republican Party introduces a resolution to remove Kevin McCarthy as Speaker of the House and a vote on the resolution is held. Both the introduction of the resolution and the vote must occur before October 1, 2023. Otherwise this question resolves as No. The outcome of the vote and any such resolutions introduced by representatives who are not Republicans are irrelevant for the purposes of this question.

\subsection*{Background Information}
Kevin McCarthy was elected Speaker of the US House of Representatives on January 7, 2023, after 15 ballots, the first time since 1923 an election for Speaker required more than one ballot. The contentious election and concessions to the House Freedom Caucus — including rules that allow any member of the House to call for a vote that would oust the Speaker by simple majority — have weakened his position as Speaker.

The rules for the House of Representatives of the 118th Congress adopt those of the 117th Congress with some amendments. The relevant portion of the rules is Rule IX, the text of which is quoted below. The amended rules remove subparagraph (3) of clause 2(a) (shown in bold).

\begin{quote}
1. Questions of privilege shall be, first, those affecting the rights of the House collectively, its safety, dignity, and the integrity of its proceedings; and second, those affecting the rights, reputation, and conduct of Members, Delegates, or the Resident Commissioner, individually, in their representative capacity only.

2. (a)(1) A resolution reported as a question of the privileges of the House, or offered from the floor by the Majority Leader or the Minority Leader as a question of the privileges of the House, or offered as privileged under clause 1, section 7, article I of the Constitution, shall have precedence of all other questions except motions to adjourn. A resolution offered from the floor by a Member, Delegate, or Resident Commissioner other than the Majority Leader or the Minority Leader as a question of the privileges of the House shall have precedence of all other questions except motions to adjourn only at a time or place, designated by the Speaker, in the legislative schedule within two legislative days after the day on which the proponent announces to the House an intention to offer the resolution and the form of the resolution. Oral announcement of the form of the resolution may be dispensed with by unanimous consent.

(2) The time allotted for debate on a resolution offered from the floor as a question of the privileges of the House shall be equally divided between (A) the proponent of the resolution, and (B) the Majority Leader, the Minority Leader, or a designee, as determined by the Speaker.

\textbf{(3) A resolution causing a vacancy in the Office of Speaker shall not be privileged except if offered by direction of a party caucus or conference.}

(b) A question of personal privilege shall have precedence of all other questions except motions to adjourn.
\end{quote}

On September 19, 2023, journalist Matt Laslo claimed to have discovered a draft motion to vacate the office of Speaker of the House in a bathroom in the US Capitol, with Matt Gaetz as the member to submit the resolution.

\end{document}